\documentclass[prl,twocolumn,showpacs]{revtex4}
\usepackage{graphicx}
\usepackage{amsmath}
\newcommand{\n}{\nonumber}
\newcommand{\bn}{\begin{eqnarray}}
\newcommand{\en}{\end{eqnarray}}
\newcommand{\eml}{\end{multline}}
\newcommand{\bml}{\begin{multline}}

\begin{document}

\title {An atomtronics transistor for quantum gates }
\author{Miroslav Gajdacz$^{1,2}$, Tom\'{a}\v{s} Opatrn\'{y}$^2$, and Kunal K. Das$^3$ }
 \affiliation{
  $^1$Institute for Physics and Astronomy, Aarhus University, Ny Munkegade 120,
8000 Aarhus C, Denmark\\
 $^2$Optics Department, Faculty of Science, Palack\'{y} University, 17.  Listopadu 12,
 77146 Olomouc, Czech Republic\\
 $^3$Department of Physical Sciences, Kutztown University of Pennsylvania, Kutztown, Pennsylvania 19530, USA}

\date{\today }
\begin{abstract}
We present a mechanism for quantum gates where the qubits are encoded in the population distribution of two component ultracold atoms trapped in a species-selective triple-well potential. The gate operation is a specific application of a new design for an atomtronics transistor where inter-species interaction is used to control transport, and can be realized with either individual atoms or aggregates like Bose-Einstein condensates (BEC). We demonstrate the operational principle with a static external potential, and show feasible implementation with a smooth dynamical potential.
\end{abstract}

\date{\today }
\pacs{03.67.Lx, 67.85.-d, 05.60.Gg, 03.75.Mn}
 \maketitle

As research in ultracold atoms shifts more towards practical applications, two of the most promising areas are that of quantum computation \cite{RMP-Delgado-qc} and atomtronics or electronics with trapped ultracold atoms \cite{Holland-PRL-atomtronics,Holland-PRA-atomtronics,Das-Aubin-PRL2009}. Whereas classical computation is intimately tied to electronics, quantum computation in the context of ultracold atoms has so far evolved independently of the relatively new field of atomtronics. Taking a cue from classical computers, it is likely that analogs of standard electronic components like diodes and transistors can be valuable in quantum computation as well. In this paper we propose a different design for an atomtronics transistor which can also be used to implement a novel mechanism for a two qubit quantum gate. Our proposal has several distinguishing features that can be advantageous: (a) qubit encoding in the spatial coordinates of particles allows implementation with single particle as well as with multi-particle entities such as BEC, (b) a system-independent principle that offers broad choices for physical realization, (c) operation does not require manipulation of internal states, and (d) easily optimizable for high fidelity and speed.

A distinguishing feature of our gate mechanism is that qubits are directly encoded in and readout from the spatial distribution of atoms.  Spatial mode encoding has been primarily used in optical qubits, in the context of continuous variable quantum computation \cite{RMP-continuous-variable}, or in dual-rail schemes \cite{Fiurasek}. Certain clever proposals for realizing phase gates in double-well potentials \cite{Teisinga-motional-qg,Mompart-gate-PRL,Clark-PRA-phase-gate} have also employed vibrational modes of trapped atoms. However, the gate outcome is contained in the phase of the states and readout was shown to require intermediate encoding on internal states \cite{Schmiedmayer-motional-qg}. In contrast, we encode qubits in the population distribution of atoms in a triple-well potential, and readout simply involves determining the presence or absence of atoms in specific wells, possible even for single atoms by direct imaging methods  \cite{Bloch-single-atom-Nature}.

Likewise, a simpler operational principle underlies our atomtronics transistor, where the atom transport is directly controlled by interspecies interaction. Existing designs are based on manipulating resonant coupling of lattice sites by adjusting the chemical potential or external bias fields \cite{Holland-PRL-atomtronics,Holland-PRA-atomtronics,Zozulya-transistor-PRA}; or on manipulating atomic internal states to transport holes \cite{Mompart-holes-PRA}, or spin \cite{Clark-spintronics-PRL-2008}.

\begin{figure}[t]
\includegraphics*[width=\columnwidth]{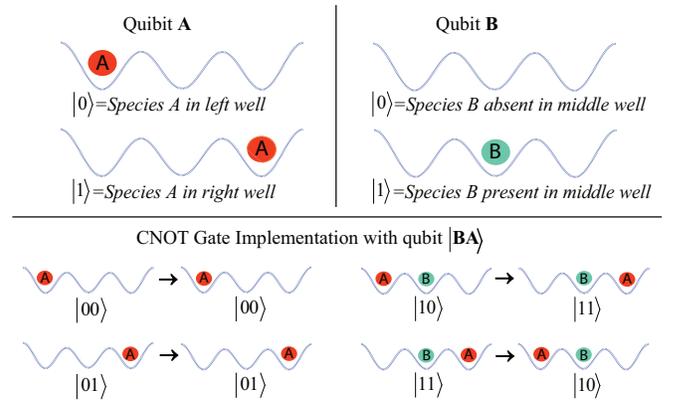}
\caption{Qubit definitions and CNOT quantum gate operation.}\label{gate}
\end{figure}

\begin{figure}[t]
\includegraphics*[width=\columnwidth]{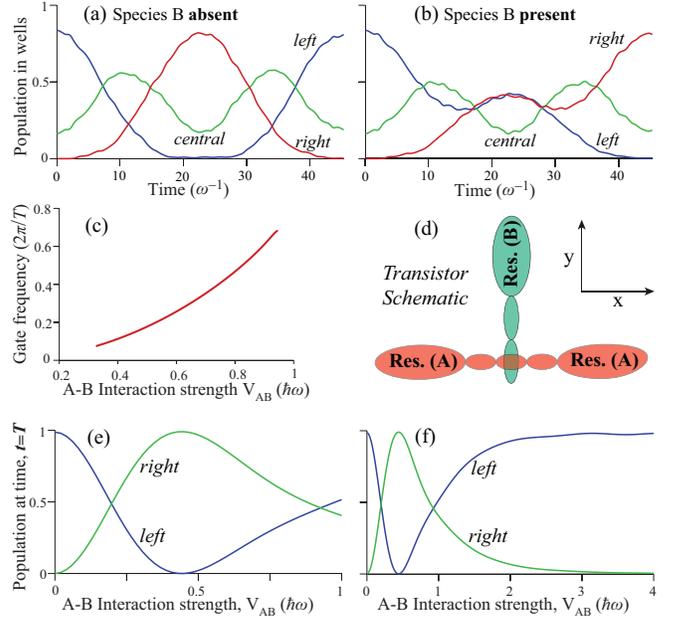}
\caption{{\bf Static potential}: Evolution of species A in a gate cycle with (a) species B absent and (b) species B present. (c) The gate frequency $2\pi/T=\Delta E_1/\hbar$ varies smoothly with $V_{AB}$.
(d) Schematic of transistor, with the central well coupled to reservoir of species B, and the extreme wells to reservoirs of species A. Transistor Operation: (e) Populations in left and right wells after one period $t=T$, as $V_{AB}$ is varied, and (f) the same showing blockade at stronger interaction.
}\label{static}
\end{figure}

{\bf Quantum Gate Operation:} We consider a triple-well potential (Fig.~\ref{gate}), with wells labeled \emph{left, central, right}, that can be populated with two \emph{repulsively} interacting species A and B. Single qubits are encoded in their spatial degrees of freedom: Qubit A is in state $|0\rangle$ or $|1\rangle$ if species A is localized in the left well or the right well respectively; Qubit B is in state $|0\rangle$ or $|1\rangle$ when species B is absent or present in the central well, with `absent' corresponding to a definite state in an adjacent transversely coupled well (see Fig.~\ref{static}(d)).
Our initial objective is to create a two-qubit CNOT gate, in a \emph{static} triple-well, such that after a set time  T, qubit A is negated if qubit B is in  $|1\rangle$, but is unchanged if qubit B is in $|0\rangle$.

\begin{figure}[t]
\includegraphics*[width=\columnwidth]{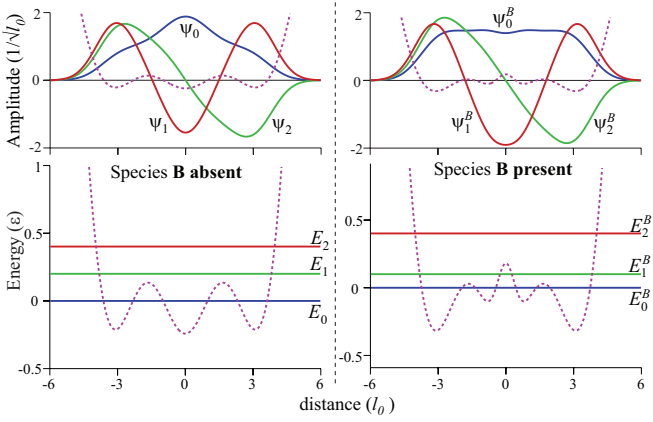}
\caption{The three lowest eigenstates (upper panels) and corresponding eigenenergies (lower panels) of species A in the triple-well, with species B absent (left) and present (right). The potential is shown with dotted lines. With B absent, $\Delta E_2=2\times\Delta E_1$ but with B present, $\Delta E_2^B=4\times\Delta E_1^B$. The energy of the ground state is always the energy reference.}\label{energies-states}
\end{figure}

Implementation can be understood in terms of the three lowest eigenstates $\phi_0,\phi_1$ and $\phi_2$ for species A in the triple-well, with eigenenergies  $E_0<E_1<E_2$. As shown in Fig.~\ref{energies-states}, the state $\phi_1$ has its \emph{node} in the central well while the other two $\phi_0$ and $\phi_2$  have anti-nodes. Therefore, when species B is present in the central well, the repulsive A-B interaction $V_{AB}$ will shift up the energies $E_0$ and $E_2$, but hardly affect $E_1$.  A class of potentials exists where the presence of atom B will raise $E_0$ and $E_2$ by the same amount, thus leaving $\Delta E_2 = E_2 - E_0$ unchanged while decreasing $\Delta E_1 = E_1 - E_0$.

Species A is prepared in a state $|\psi_{A}(t=0)\rangle$ localized in one of the two extreme wells. By choosing it to be a Gaussian with minimized energy for that well, we ensure almost complete projection on the three lowest eigenstates. The initial phase relations among the eigenstates, shown in Fig.~\ref{energies-states}, are such that $\phi_0(0)$ and $\phi_2(0)$ add up constructively with $\phi_1(0)$ in one extreme well and destructively in the other. If present, $|\psi_{B}(0)\rangle$ is a minimum energy Gaussian localized in the central well, and we assume a factorized initial two-particle state $|\psi_{AB}(0)\rangle=|\psi_{A}(0)\rangle|\psi_{B}(0)\rangle$ (justified at the end of the paper). The process is made to work starting from either extreme well by always keeping the potential bilaterally symmetric about the central well minimum. We adjust \emph{four} degrees of freedom of the system: position, width and height of the internal barriers, and $V_{AB}$. By simple reparametrization, two are used to fix a time scale of operation, and to set the condition that with species B \emph{absent}, the energy differences satisfy $\Delta E_2=2\times\Delta E_1$ so that after time $T=h/\Delta E_1$, the dynamical phase acquired by the three eigenstates are offset by multiples of $2\pi$ leading to the revival of the initial state in the \emph{initially occupied} extreme well. The remaining two parameters are set to ensure that with species B \emph{present}  $\Delta E_2^B=\Delta E_2$ remains unaltered while $\Delta E_1^B=\Delta E_1/2$ is halved, as seen in Fig.~\ref{energies-states}, so that now $\Delta E_2^B=4\times\Delta E_1^B$. The antisymmetric state $\phi_1$ evolves at half the rate than without B, hence after the same time of evolution T it has an opposite phase, or $\pi$ offset, relative to the symmetric states. This results in localization of the species A in the \emph{initially empty} extreme well.

Thus, the presence or absence of the species B leads to the revival of species A in the original well or transfer to the other extreme well after a set time $T$, implementing the CNOT gate as shown in Fig.~\ref{gate}.  The evolution of the population in the three wells during a gate cycle is shown in Fig.~\ref{static} (a) and (b) when species A is initially localized in the left well. The populations in the wells are computed by integrating $|\psi_{A}(x,t)|^2$ over the intervals $(\infty,-D),(-D,D),(D,\infty)$, where $\pm D$
are the coordinate of the barrier maxima.  Figure~\ref{static} (c) shows the range of gate frequencies ($2\pi/T$) for which the gate conditions ($\Delta E_2=2\times\Delta E_1$, $\Delta E_2^B=\Delta E_2$ and
$\Delta E_1^B=\Delta E_1/2$) can be fulfilled and the corresponding interaction strength.

{\bf Transistor Operation:} The gate operation described above is the specific case, of fixed $V_{AB}$, of a general transistor mechanism in which species B is used to precisely control the flow of species A among the extreme wells. A schematic of operation as an atomtronics transistor is shown in Fig.~\ref{static} (d), where the left and right wells of the triple-well setup are coupled to two reservoirs of species A, while the central well is coupled transversely via an adjacent well to a reservoir for species B. Since  $V_{AB}$ directly affects the dynamic evolution of species A it controls the transfer rate from the left well to the right well.  Figure~\ref{static} (c) shows that a range of interaction strengths are available to smoothly adjust $\Delta E_1$ which fixes the period for a gate cycle $T=h/\Delta E_1$.  Therefore, if we instead fix the period of each cycle at T and vary $V_{AB}$, the  transfer per cycle can be controlled; Fig.~\ref{static} (e) shows that the fraction of species A transferred varies smoothly from zero to complete transfer. Since $V_{AB}$ depends on the interaction strength $g_{AB}$ and the density $|\psi_B|^2$ of species B, for large $g_{AB}$ small variations in $|\psi_B|^2$ can be used to control a large flux of species A, creating an amplification effect. Eventually, as $V_{AB}$ is further increased Fig.~\ref{static}(f) shows that transfer of species A is completely blocked.

{\bf Implementation with trapped atoms:} We present simulations for ultracold atoms in quasi-one-dimensional potentials \cite{das-crossover} where only the transverse ground state is populated. Species selective potential \cite{Hansch-collisional-qg} create different directions of relaxed confinement for the two species, $x$ for A and $y$ for B, as shown in Fig.~\ref{static}(d). This keeps species B localized (when present) in the central well of the triple-well potential along the x-direction, but it is free to move in the y-direction, so its presence in the central well can be varied. Our analysis can therefore be restricted to the dynamics in the x-direction, along which we assume harmonic potential for the outer shape of the triple-well (excluding reservoirs); it's angular frequency $\omega$ sets our units: energy $\hbar\omega$, length $l_0=\sqrt{\hbar/(m\omega)}$ and time $\omega^{-1}$.  The triple-well is completed with two identical Gaussian profile barriers symmetrically placed at distance, $d$ from the harmonic potential minimum (our coordinate origin). For both species the net external potential along the x-direction has the same form:
\bn V_A(x,t)&=&{\textstyle\frac{1}{2}}x^2
+
U[e^{-\frac{(x-d)^2}{2\sigma^2}}+e^{-\frac{(x+d)^2}{2\sigma^2}}].\label{potential}\en
and $V_B(x,t)=20\times V_A(x,t)$. In the \emph{static} model presented above $V_i(x,t)=V_i(x)$ is unchanged over time, but in the dynamical model we present later, the strength $U(t)$ and the position $d(t)$ of the barriers evolve in time.

The simplest feasible model involves a single atom of each species A and B, with effective one-dimensional hard-core bosonic interaction strength $g_{AB}$:
\begin{equation}
\hat{H}=\sum_{i={A,B}}\left(-\frac{1}{2}
\frac{\partial^{2}}{\partial x_{i}^{2}}+V_i(x_i,t)\right)+g_{AB} \delta(x_A-x_B)\label{hamiltonian}
\end{equation}
obtained from the 3D Hamiltonian for hard core bosons by integrating out the transverse degrees of freedom \cite{das-crossover}.

The dynamics of species A has little effect on that of species B due to its strong confinement, so we may factorize the two-atom wavefunction $|\psi_{AB}\rangle=|\psi_A\rangle|\psi_B\rangle$. By taking the projections $\langle\psi_B|H|\psi_{AB}\rangle$ and $\langle\psi_A|H|\psi_{AB}\rangle$, the equations of motion for the Hamiltonian can be reduced to two coupled equations:
\bn -\frac{1}{2}\partial_{xx}\psi_A+V_A\psi_A+g_{AB}|\psi_B|^2\psi_A=i\partial_t\psi_A\n\\
-\frac{1}{2}\partial_{xx}\psi_B+V_B\psi_B+g_{AB}|\psi_A|^2\psi_B=i\partial_t\psi_B.\label{BEC hamiltonian}\en
The assumption of strong $V_B$ also implies that the interaction term $g_{AB}|\psi_A|^2\psi_B$ has little effect on the evolution of $\psi_B$ as we verified. These equations can represent two coupled BEC's; self interactions $g_{A(B)}|\psi_{A(B)}|^2$ can be added, but not considered in this paper. We use \emph{both} scenarios for all our results presented here: (i) the two particle Hamiltonian in Eq.~(\ref{hamiltonian}) and (ii) the coupled equations in Eq.~(\ref{BEC hamiltonian}) and find them almost indistinguishable.

\begin{figure}[t]
\includegraphics*[width=\columnwidth]{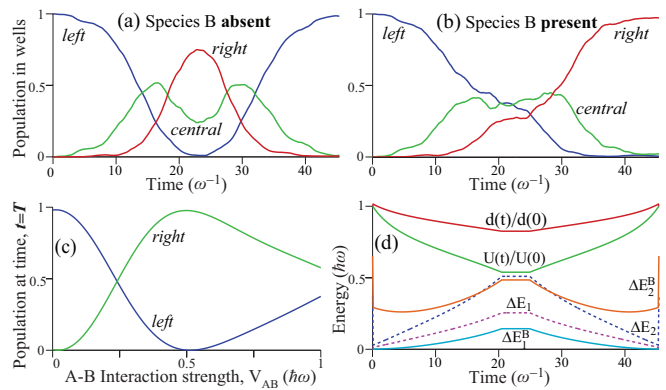}
\caption{{\bf Dynamic Potential}: Evolution of species A in a gate cycle with (a) species B absent and (b)  species B present. (c)  Populations in left and right wells after one period $t=T$, as $V_{AB}$ is varied. (b) Smooth time evolution of the dynamical parameters: barrier position $d$  and barrier amplitude $U$; also shown are the energy separations with species B absent ($\Delta E_1^B,\Delta E_2^B$, dotted lines) and present ($\Delta E_1^B,\Delta E_2^B$).}\label{dynamic}
\end{figure}

{\bf Dynamic Gate and Transistor:} The model presented so far assumed
a static potential that does not vary during operation,
and underscores the fact that both gate and transistor mechanism operate by evolution of the dynamical phases of the states. However, to put this in practice, two issues need to be addressed: (i)
preparation and readout of the quantum bit and (ii) initiation and termination of each cycle of operation.
Both goals can be achieved, provided that the initial and final states of species A are the ground state of a single isolated well. The population would stay localized in the extreme well as long as the barriers are kept high, and the spatial isolation enables adiabatic loading and readout of the state.

The gate operation is turned on smoothly by decreasing barrier height $U(t)$ and separation $2d(t)$. Being already a superposition of the three lowest eigenstates, $|\psi_{A}(t)\rangle$ adiabatically follows their evolution, and the evolution of their relative phases performs the gate operation as before. Termination mirrors initiation, since the second half of the process is the reverse of the first half as shown in Fig.~\ref{dynamic}(d).

\emph{Transport algorithm:}  For simplicity, the path selection is based only on the properties of the potential without species B; in its presence optimum transfer of species A is achieved by a proper choice of the interspecies interaction. The two  parameters $U(t)$ and $d(t)$ are used to adjust
$\Delta E_1$ and $\Delta E_2$. The barrier width $\sigma$ is kept constant here, but could  be varied in time as an additional control parameter. We start from a potential with high barriers and slightly raised bottom of the middle well (with tunneling completely suppressed) we rapidly put the three wells into degeneracy, setting $\Delta E_2 = 2 \times \Delta E_1 = 0.28 \hbar\omega$ by a tiny adjustment in $d(t)$, barely visible in Fig.~\ref{dynamic}(d). This step demonstrates the physical isolation of the wells achievable at the start and the end of a cycle.

While maintaining  $\Delta E_2 = 2 \times \Delta E_1$, we slowly increase $\Delta E_1$ and monitor the phase evolution of the eigenstates $\phi_i$. The speed of variation of the potential
is limited by the non-adiabatic coupling of the three lowest states to the higher order states, defined to be:
\bn
{\cal A}_{kn}=i\hbar\frac{\langle \phi_k|\partial_t\hat{H}|\phi_n\rangle}{(E_k-E_n)^2}.
\en
Due to symmetry of the Hamiltonian, coupling occurs only between states of the same parity.
The coupling ${\cal A}_{02}$, is irrelevant since it simply splits ${E_0}$ and ${E_2}$ further apart maintaining $\Delta E_2= 2\times\Delta E_1$. So, we ensure adiabatic operation by
setting the next higher coupling ${\cal A}_{13} = 0.07$, a low value. The path selecting algorithm stops when the dynamical phase $t\Delta E_1=\pi$ (half of the revival); the second half is obtained by time reversal.

We ensure that with species B present, exactly the same path leads to the desired outcome, of species A localizing in the opposite well, by tuning $V_{AB}$ to maximize the fidelity. The fidelity is measured by the probability of the projection of the final state at $t=T$ on the desired CNOT gate outcomes shown in Fig.~\ref{gate}, and it directly correlates to the transferred fraction of species A after one cycle, plotted in Fig.~\ref{dynamic}(c). For both static and dynamic potentials, the fidelity was about $98\%$.

\begin{figure}[t]
\includegraphics*[width=\columnwidth]{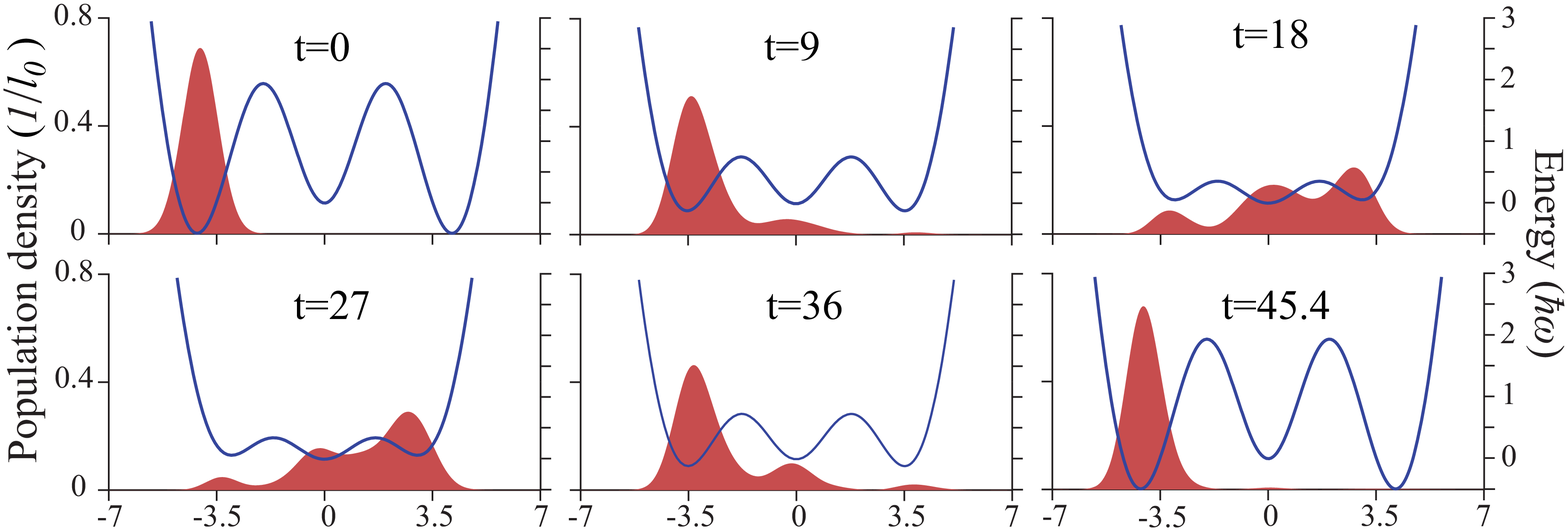}
\includegraphics*[width=\columnwidth]{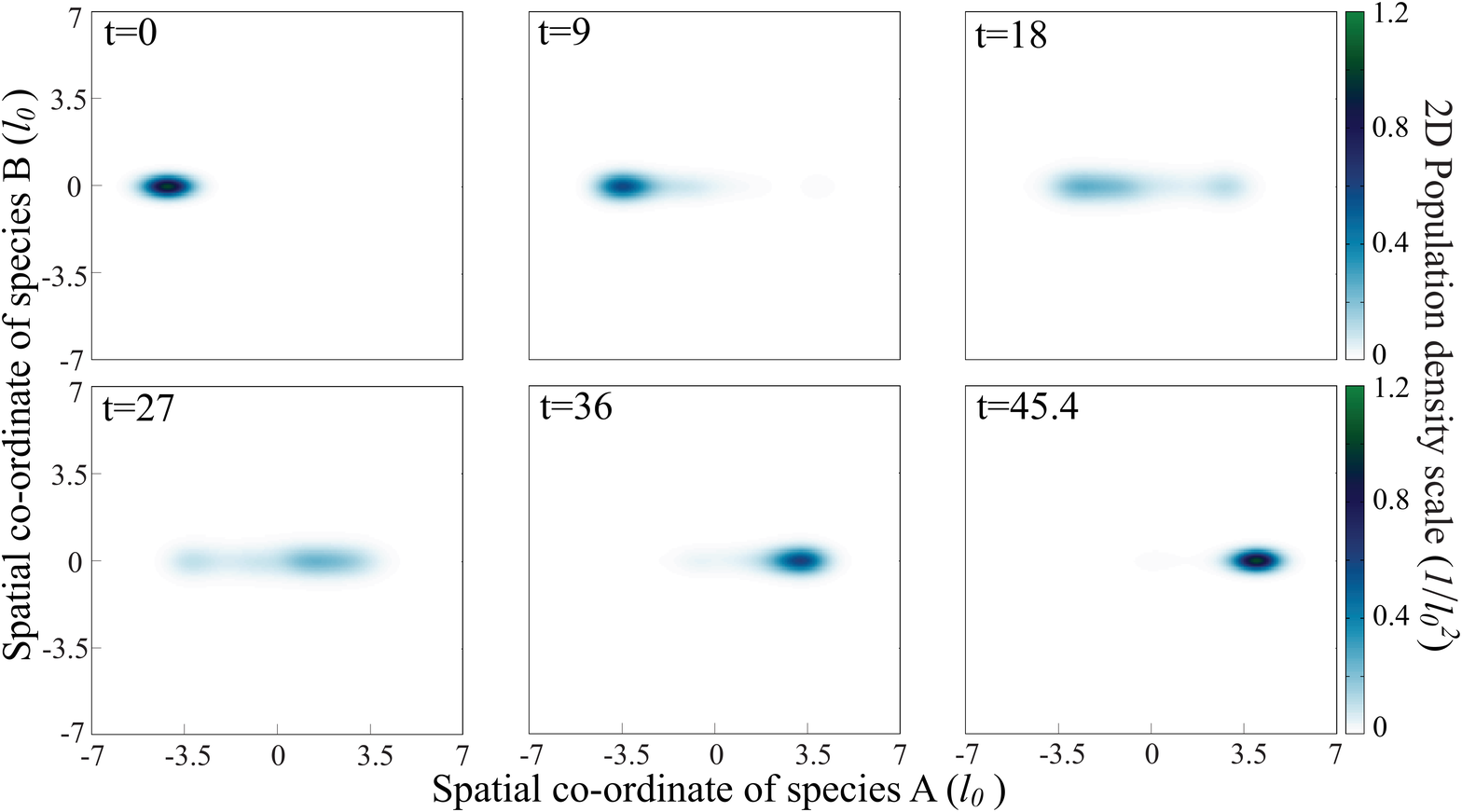}
\caption{Snapshots \cite{movies} of evolution of population density in a gate cycle of period T (time units of $\omega^{-1}$) corresponding to Fig.~4(a) and 4(b): \emph{upper panels}: Density (left axis) of species A when species B is absent; the potential is shown as blue lines (right axis); \emph{lower panels}: Joint density of species A (horizontal axis) and species B (vertical axis) when species B is present. The two-particle Hamiltonian in Eq.~(\ref{hamiltonian}) is used here; evolution is identical with the coupled equations Eq.~(\ref{BEC hamiltonian}).}\label{snapshots}
\end{figure}

{\bf Discussion and Conclusions:} Snapshots \cite{movies} of the dynamics for gate operation shown in Fig.~\ref{snapshots} confirm that: (i) species B remains localized and is not influenced by the evolution of species A, (ii) species A is significantly delocalized during transit with presence in the central well
(iii) dynamics with Eq.~\ref{hamiltonian}, and Eq.~\ref{BEC hamiltonian} were indistinguishable on the scale of the plots, suggesting implementation with both individual atoms and BECs. With BEC, Feshbach resonance can be used to suppress self-interaction.  With individual atoms optical lattices may be preferable and our results would apply with a simple reparametrization \cite{GOD-1}. The parameters in our simulations are  $U=13.3\hbar\omega, \sigma=2.35 l_0, d=2.08 l_0$ (static) and $U(0)=12.9\hbar\omega, \sigma=1.61 l_0, d(0)=1.82 l_0$ (dynamic) and $g_{AB}=0.44\ \hbar\omega l_0$. Due to proximity of the barriers in Eq.~(\ref{potential}), the well depths (barrier-peak to well-bottom) never exceed $\sim 3\hbar\omega$.  On an atom chip \cite{Hansch-collisional-qg} with $\omega\simeq 2\pi\times5$ kHz , the parameters used in our simulation correspond to  $T\simeq 1$ ms  for a gate cycle; similar timescale is obtained when we re-parameterize for a lattice.

After a gate cycle, the reduced density matrix $\rho_A(t)=\rm{Tr}_B|\psi_{AB}(t)\rangle\langle\psi_{AB}(t)|$, gives $1-{\rm Tr}\rho_A^2(T)=4.4\times10^{-4}$, indicating a pure state. The von Neumann entropy $S(T)=-{\rm Tr}[\rho_A(T) \ln(\rho_A(T)]=2.3\times 10^{-3}$, with $S(t)<4.0\times 10^{-3}$ during the cycle for the dynamic potential and $S(t)<12\times 10^{-3}$ for the static. These justify the factorization of $|\psi_{AB}\rangle$ and explains the agreement between using Eq.~\ref{hamiltonian} and Eq.~\ref{BEC hamiltonian}; they also show absence of significant entanglement between the species. Gate fidelity can be arbitrarily improved with optimal control methods \cite{Sklarz-PRA-2002} since the mechanism has well-defined initial and target states and a highly optimal initial path; but we leave out such results as they are system specific. This design can implement a universal set of gates, since single qubit gates, such as a Hadamard gate, can also be implemented by adjusting  $\Delta E_{1}$, with species B absent, for desired partial transfer in a cycle.

T.O. and M.G. acknowledge support of the Czech Science
Foundation, Grant No. P205/10/1657. K.K.D. acknowledges
support from NSF  Grant No. PHY-0970012.

\end{document}